\documentclass[3p]{elsarticle}

\usepackage{xr}
\usepackage{graphicx, amsmath, amssymb, amsfonts, mathtools, mathrsfs, color}
\usepackage[export]{adjustbox}
\usepackage{comment, enumerate, tabularx, multirow}
\usepackage{natbib, hyperref, url}
\usepackage{algorithm, algpseudocode, pifont, longtable}

\newcommand{\abs}[1]{ \left| #1 \right| }
\newcommand{\sgn}{\operatorname{sgn}}
\newcommand{\DiffLim}{\alpha_{\text{lim}}}
\newcommand{\Tavg}{T_0}
\newcommand{\rhoavg}{\rho_{0}}
\newcommand{\cp}{c_{\text p}}
\newcommand{\cv}{c_{\text v}}
\newcommand{\Nu}{\text{Nu}}
\newcommand{\Ra}{\text{Ra}}
\newcommand{\cw}{\chi}
\newcommand{\Twavg}{{T}_{wa}}

\newcommand{\CO}{CO$_2$}
\newcommand{\source}{\Phi}

\begin{document}
\title{A minimal model for predicting ventilation rates of subterranean caves}

\author[GFDI]{Karina Khazmutdinova}
\author[GFDI,EOAS]{Doron Nof}
\author[Oberlin]{Darrel Tremaine}
\author[GFDI,EOAS]{Ming Ye}
\author[GFDI,Math]{M.~N.~J.~Moore}

\address[GFDI]{Geophysical Fluid Dynamics Institute, Florida State University, USA.}
\address[EOAS]{Department of Earth, Ocean, and Atmospheric Science Department, Florida State University, USA.}
\address[Oberlin]{Office of Environmental Sustainability, Oberlin College, USA}
\address[Math]{Department of Mathematics, Florida State University, USA.}






%
%
\begin{abstract}
The ventilation of carbon dioxide within subterranean caves regulates the growth
of speleothems --- mineral deposits found in caves that provide important clues about past climate. While previous studies have used internal temperature measurements to predict ventilation rates, such data would not be available for the task of climate reconstruction. Here, we develop a parsimonious model to predict ventilation rates from knowledge of outside temperatures and the cave's physical dimensions only. In the model, ventilation arises from buoyancy-driven flows created in passageways that connect to the outside. A few key simplifications leads to a system amenable to perturbation analysis, resulting in explicit expressions for ventilation rates. We compare these predictions to time-resolved, in-situ measurements of transported cave gases (carbon dioxide and radon). The theory accurately accounts for the timing and magnitude of seasonal and synoptic variations of these gases, and is therefore diagnostic of seasonal bias in speleothem growth.
\end{abstract}
\maketitle

\section{Introduction}
Subterranean caves contain calcium-carbonate ($CaCO_3$) mineral deposits, known as speleothems, which record important clues about past climate. Much like tree-rings and ice cores, these formations can be used to infer past conditions and identify major climatic shifts \citep{bar1997late, cruz2007evidence, wang2008millennial}. Speleothem records, however, can be difficult to interpret due to a number of competing factors that influence chemical and isotopic composition, mineral fabric, and timing of deposition \citep{frisia2000calcite, fairchild2006modification, lachniet2009climatic, fairchild2012speleothem, wong2015advancements}. Early cave researchers realized the potential importance of internal cave conditions \citep{broecker1960radiocarbon, holland1964some, hendy1971isotopic}, but only within the past twenty years has \textit{in situ} monitoring technology become sufficiently advanced to record microclimate data with which to calibrate speleothem records \citep{collister2005high, perrier2005modelling, spotl2005cave, baldini2006spatial, banner2007seasonal, mattey2010seasonal, kowalczk2010cave, luetscher2012cora, baker2014reconstructing, tremaine2015dynamic}. These studies highlighted the fact that in order to fully understand a speleothem paleoclimate record, it is necessary to have a holistic understanding of the site-specific parameters that control stalagmite growth. These include but are not limited to rainfall amount, hydrologic saturation, vegetation and soil productivity, dripwater residence time and water-rock interaction, drip rates, and ventilation \citep{fairchild2012speleothem}.  

Speleothems are the end product of a series of chemical reactions: (1) rainwater combines with soil-zone carbon dioxide to form carbonic acid: $H_2O + CO_{2 (g)} \to H_2CO_3$; (2) carbonic acid percolates downward and slowly dissolves carbonate bedrock, creating a high pCO$_2$ solution of calcium and bicarbonate: $H_2CO3 + CaCO_3 \to Ca^{2+}_{(aq)} + 2HCO_{3 (aq)}^{-}$: and (3) the drip reaches a cave void, where the gradient between low-pCO$_2$ cave air and high-pCO$_2$ dripwater causes the drip to degas CO$_2$, driving the pH of the solution up and causing re-precipitation of calcium carbonate: $Ca^{2+}_{(aq)} + 2HCO_{3 (aq)}^{-} \to CO_{2 (g)} + H_2O + CaCO_3$ \citep{plummer1982solubilities}. Cave ventilation modulates cave air CO$_2$ concentration and thus the gradient between air and drip, driving both the timing and the vigor of speleothem growth. Without ventilation, cave air CO$_2$ would come to chemical equilibrium with dripwater CO$_2$, and speleothem formation would cease. 

In addition to controlling the rate and timing of deposition, ventilation has been shown in several long-term cave monitoring studies to have significant impacts on the isotopic composition of dripwater and subsequent speleothem calcite \citep{spotl2005cave, frisia2011carbon, lambert2011controls, tremaine2011speleothem, feng2012oxygen}. Results from other field, theoretical, and laboratory studies support the hypothesis that under strong ventilation regimes where CO$_2$ is quickly removed from dripwater, rapid precipitation may induce kinetic isotope fractionation, causing a shift in oxygen and carbon isotopes within the speleothem and making paleoclimatic interpretation of the record much more difficult \citep{wiedner2008investigation, scholz2009modelling, muhlinghaus2009modelling, polag2010stable, dreybrodt2011climatic, stoll2015interpretation}.

Because ventilation controls the rate of calcite precipitation, it not only influences \textit{when} a speleothem forms (which seasons), but also \textit{where} the speleothem forms; i.e. on or within the cave ceiling (stalactite) or on the cave floor (stalagmite). Wherever dripwater encounters low-CO$_2$ cave air, there is potential for calcite precipitation up-stream of the stalagmite. This phenomenon is known as Prior-Calcite-Precipitation, or PCP \citep{fairchild2000controls}. When PCP occurs, dripwater cation-to-calcium (X/Ca) ratios change as Ca is preferentially removed from solution and minor elements become more concentrated \citep{fairchild2012speleothem}. While PCP is often associated with climate-driven water balance above the cave, where more/less hydrologic saturation and rainfall amount above the cave results in higher/lower drip rates, lower/higher PCP, and lower/higher speleothem X/Ca ratios \citep{johnson2006seasonal, karmann2007climate, mcdonald2007hydrochemical, verheyden2008monitoring, tremaine2013speleothem, khazmutdinovapercolation, tadros2016enso}, strong ventilation and low cave-air CO$_2$ also promotes PCP and exerts control on variations in speleothem x/Ca ratios \citep{fairchild2009trace, sherwin2011cave, wong2011seasonal, treble2015impacts}.

Given that the process of cave ventilation can have significant impacts on the isotopic and trace element composition of modern calcite, we must endeavor to understand the ventilation of each cave that contains speleothems of paleoclimatic importance. This is especially true in light of recent work that recognized how cave ventilation varies seasonally with well-defined global patterns \citep{Hardt2015}. While past studies have quantified cave ventilation in terms of internal temperature measurements \citep{de1987cave, christoforou1996air, gregorivc2014role}, such data would not be available for the task of climate reconstruction. Instead, predictive models must be developed.

Here, we construct a theoretical framework to predict ventilation rates from a minimal set of external information. Building on previous work \citep{wigley1971geophysical, de1987cave, christoforou1996air, kowalczk2010cave}, we model ventilation as arising from buoyancy-driven flows created by the internal-external temperature difference. \cite{faimon2012air} showed that the temperature explained more than 99\% of variations in cave air density, thus, we used temperature difference to model ventilation rates. Unlike previous studies, though, our model does not rely on any internal measurements. It only requires basic physical parameters of the cave itself and easily obtainable outside temperature data. To close the model, we require the flow to be critical via the composite Froude number --- a condition borrowed from the field of oceanography \citep{armi_1986, armi1986maximal, dalziel1991two, helfrich1995time, pratt2008critical}. Analysis of the relevant scales, along with a few key simplifications, produces a system that can be solved through perturbation methods, ultimately giving explicit formulas for how ventilation depends on the system's physical parameters.

To test our ventilation model, we use two tracers common to all limestone caves; carbon dioxide (CO$_2$) and radon-222 ($^{222}$Rn). As previously discussed, cave air CO$_2$ is derived from microbial decay and plant-root respiration in the soil zone, as well as dissolved limestone bedrock. Cave air CO$_2$ concentrations are modulated by a combination of input from both dripwaters and gaseous transport through cracks and fissures above the cave, import of atmospheric CO$_2$ via ventilation, and removal via condensation corrosion \citep{sanchez1999inorganic, gabrovvsek2000role, baldini2006carbon, mattey2016carbon}. $^{222}$Rn is a daughter isotope in a radioactive decay chain that begins with $^{238}$U and ends with $^{206}$Pb. Limestone contains an average 1.3-2.5ppm $^{238}$U, and therefore acts as a continuous source of radon production and advection into cave chambers through dripwaters, surfaces, and cracks \citep{fernandez1986natural, hakl1997radon}. $^{222}$Rn is an ideal tracer for cave ventilation because atmospheric concentrations of radon are negligible and it is an inert gas with a 3.82-day half-life, which is sufficiently long when compared with most cave-air exchange rates \citep{fernandez1986natural, perrier2005modelling, richon2004spatial, kowalczk2010cave}. We report a set of time-resolved measurements of cave air gas concentrations from a subterranean cave located in Florida Caverns State Park \citep{khazmutdinova_cave}. The theory accounts for qualitative trends in the measurements throughout the year, and, over certain periods, even shows quantitative agreement with weekly-scale fluctuations. This comparison also allows the production rates of CO$_2$ and $^{222}$Rn to be estimated, which could complement other, more invasive, estimation techniques \citep{faimon2006anthropogenic, kowalczk2010cave}.
\section{Study site and micrometeorological station}
Dragon's Tooth Cave (DTC) is a pristine, karst-hosted dry cave located in Florida Caverns State Park, Marianna, northwest Florida (Fig. 1a-b). The cave formed within a carbonate bluff along the nearby Chipola River, and has two primary passages that trend northwest to southeast in alignment with regional bedrock fractures. The two large passages are intersected by a third east-west trending corridor that dips at approximately 45 degrees from west to east, serving as an entrance to both sections. The northern section was first surveyed and mapped in 1978. The floor of the northern section lies at 18 meters above sea level (ASL) and is composed of mud-clay, which is typical of caves in Marianna that are subjected to seasonal groundwater inundation when Chipola River levels exceed 3-4 meters above the riverbed \citep{tremaine2015dynamic}. Passages within DTC are 2-3 meters tall, thus floodwaters tend to fill the passages and remove any stalactites or soda straws that may have formed, minimizing the potential of flowstone decoration. 

In 1984, spelunkers observed airflow through a pile of breakdown near the entrance and dug a narrow tunnel making a connection to the southern section of the cave, finding a previously unexplored room that was highly decorated with beautiful speleothem formations. This room later came to be known as the Dragon's Belly (DB) \citep{maddox1986DTC}. The floor of the Belly lies at 21 meters ASL; groundwater slowly fills the Belly from below only during Chipola flooding events that exceed 5 meters above riverbed. Topography above the cave suggests that the Belly is the remains of a large diameter NW-to-SE conduit that has collapsed at both ends of the room. The porous nature of the breakdown materials promotes significant volumetric airflow (ventilation) and therefore the Belly is highly decorated with pristine white stalagmites, flowstones, and moon-milk atop several large sections of breakdown blocks. In 2012, our group mapped the DTC using Leica ScanStationP20 to obtain highly accurate measurements of its dimensions. The main chamber of the DB is nearly rectangular -- 41 m long, 12 m wide, and 6 m high with 45 degree slopes at each end of the room that pinch off at the ceiling, and it connects to the entrance of the cave system via a single, narrow passageway (0.6~m high, 0.3~m wide and 3~m long). These geometrically simple features make the DB an ideal site for testing basic ventilation models. Once a firm theoretical foundation is established, future studies could examine more complex cave systems (e.g. complex geometries and/or multiple openings).

\begin{figure}
\includegraphics[width=0.8\textwidth,left]{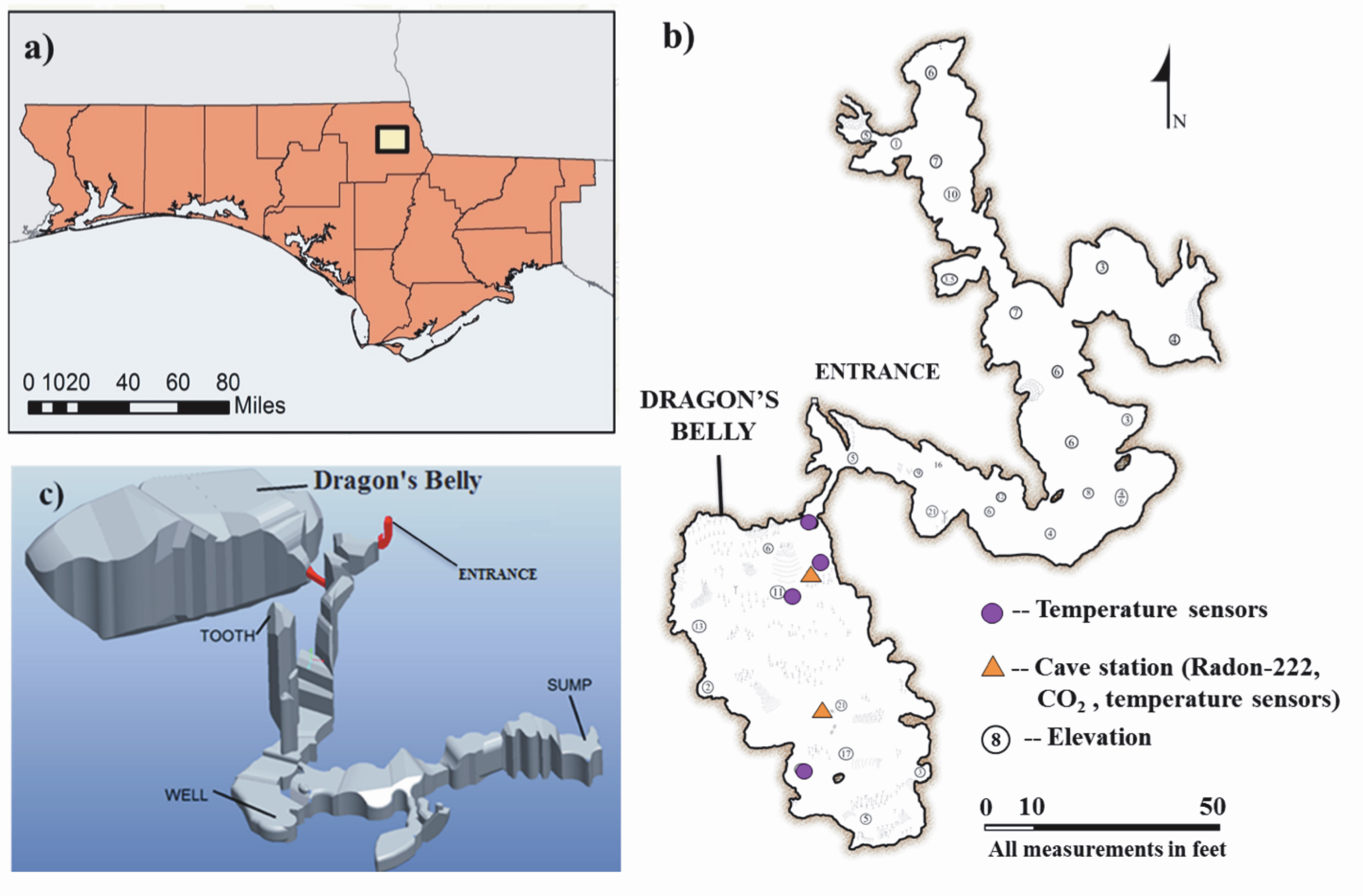}
\caption{Study site: (a) Florida Caverns State Park is located in the Florida panhandle, USA. (b) Dragon's Tooth is a system of caves which includes the main study site, the Dragon's Belly. Also shown are the locations of our sensors within the Dragon's Belly (colored symbols) and elevations (circled numbers). (c) A 3D rendering of the Dragon's Tooth Cave system shows the approximately rectangular geometry of the Dragon's Belly and the narrow passageway that connects to the entrance of the cave system (highlighted in red).}
\label{Fig1}
\end{figure}
Beginning in November 2011, we established a micrometeorological monitoring program inside the DB cave to continuously collect spatial and temporal variations of $^{222}$Rn, CO$_2$, and temperature. $^{222}$Rn and CO$_2$ levels were measured via two cave stations deployed 15 meters apart (see Fig. 1b). These stations were equipped with Durridge RAD7 detectors to measure radon, LiCor-820 gas analyzers to measure CO$_2$, as well as Vaisala HMP45C temperature and relative humidity sensors. Four additional Onset HOBO Pro v2 External Temperature Data Loggers were also deployed throughout the Belly. All data was recorded hourly, with the exception of occasional outages due to flooding events or power failures. Meanwhile, we obtained external temperature data from daily measurements taken at Marianna Municipal Airport, located 8 km from the study site.

\section{The ventilation model}
\begin{figure}
\includegraphics[width=0.8\textwidth,left]{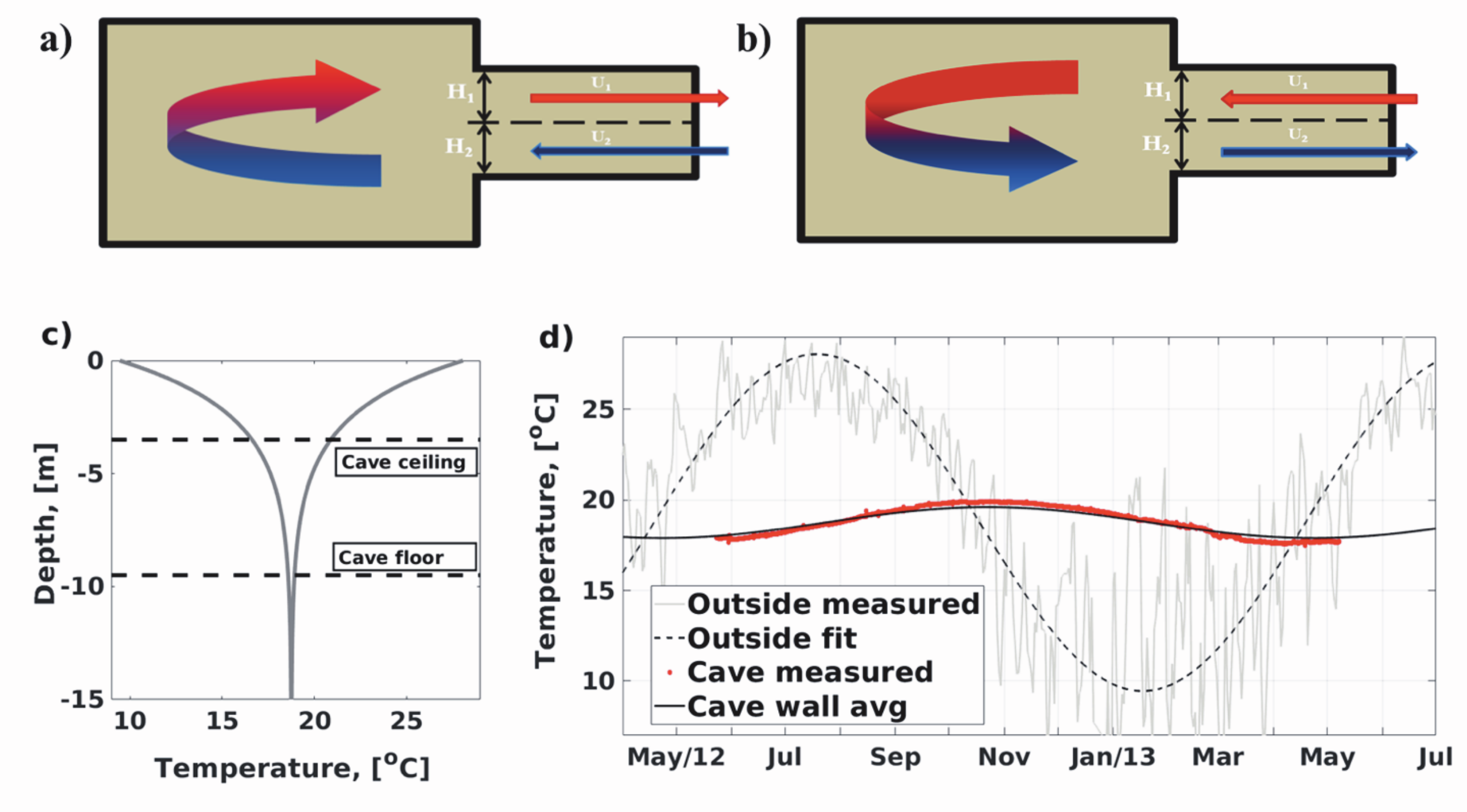}
\caption{The ventilation model: (a)--(b) The temperature difference between the cave and the outside creates an exchange flow, regardless of whether the cave is (a) warmer or (b) cooler than the outside.
(c) The temperature range at various depths as predicted by the wine-cellar formula, Eq.~(\ref{winecellar}), with the ceiling and floor of the Dragon's Belly Cave indicated.
(d) The cave-wall temperatures can be estimated by inserting the outside temperatures (measurements in faint gray and seasonal fit in dashed black) into Eq.~(\ref{winecellar}). The solid black line shows the predicted average wall temperature, which agrees well with internal measurements (red).}
\label{Fig2}
\end{figure}

We model the Dragon's Belly Cave as a well-mixed compartment with a single opening that exchanges heat with the surrounding surfaces and the outside environment. Because the internal cave temperature $T_c$ is typically different from the external temperature $T_e$, a buoyancy-driven flow is created in the connecting passageway \citep{hunt1999fluid, pratt2008critical, camassa2012}. During the summer, warm outside air flows into the cave through the opening, is cooled by contact with the cave walls, and exits through the lower part of the same opening. The airflow reverses during the winter; see Figs.~\ref{Fig2}a--b for a schematic. As long as vertical mixing and viscous effects are weak (i.e.~large Peclet and Reynolds numbers respectively), the flow in the opening is essentially that of two layers --- a top layer of depth $H_1$ and speed $U_1$, and a bottom layer with values $H_2$ and $U_2$. Enforcing conservation of volume inside the cave, along with the Boussinesq approximation, results in a {\em symmetric} exchange flow; i.e.~both the depths and speeds are the same in the two layers, $H_1 = H_2 = H$ and $U_1 = U_2 = U$. Clearly, $H$ must be the half-height of the passageway ($H = 0.3$~m), but $U$ must be determined through other considerations, starting with a balance of the cave's total thermal energy.

\subsection{Thermal energy balance and closure condition}
\label{ThermalSection}
Under the assumptions of (1) well-mixed air in the cave interior, (2) the Boussinesq approximation, and (3) a symmetric two-layer exchange flow in the cave opening, the rate of change of the cave's thermal energy takes the form
\begin{equation} 
\label{caveheatEq}
\rhoavg \cv V  \, \dot{T}_c = \rhoavg \cp A U (T_e - T_c) + \int_S q \, dS \, .
\end{equation}
The main unknowns in this ordinary differential equation (ODE) are the cave's internal temperature, $T_c$, and the exchange speed, $U$. Meanwhile, the outside temperatures, $T_e$, will be taken from the measurements at Marianna Airport. The parameters are the mean air density $\rhoavg=1.205$~kg/m$^3$, the specific heat of air at constant pressure $\cp=1005$~J/(kg~K) and at constant volume $\cv = 717$~J/(kg~K), the total volume of the cave chamber $V~=~2952$~m$^3$, and the half-area of the cave passageway $A = 0.09$~m$^2$.

The first term on the right of Eq.~(\ref{caveheatEq}) represents thermal exchange with the outside via the two-layer flow, while the second represents exchange with the cave walls through surface integration of the local heat-transfer-per-unit-area $q$~[J/(m$^2$s)]. Temperature differences between these surfaces and the enclosed air give rise to convective flows, which can promote heat transfer greatly. The Nusselt number $\Nu$ (the ratio of convective to conductive heat transfer) quantifies this increase and can be estimated through knowledge of the Rayleigh number, $\Ra=g\ell^3\abs{T_w~-~T_c}/(\nu\alpha\Tavg)$. Here, $T_w$ is the temperature and $\ell$ is a lengthscale of the surface under consideration, while $g = 9.8$~m/s$^2$ is the gravitational acceleration, $\nu=15.1 \times10^{-6}$~m$^2$/s and $\alpha=2.12 \times 10^{-5}$~m$^2$/s are the viscosity and thermal diffusivity of air, and $\Tavg = 291.9$~K is the mean annual temperature in Marianna, Florida. To estimate the Nusselt number, we use the well-established scaling law \citep{holman2002heat, bergman2011fundamentals}
\begin{equation}
\label{NuRa}
\Nu = \cw \, \Ra^{1/3} \, .
\end{equation}
Here, $\cw$ is a constant that depends on the orientation of the surface and the direction of the thermal gradient. The sidewalls produce convection regardless of the sign of the temperature gradient, whereas the floor and ceiling only produce convection in the presence of an unstable temperature gradient (see section S1 in the supporting information for further details). The above law holds for $\Ra > 10^7$, and inside the Dragon's Belly we typically have $\Ra \sim 10^{10}$. 

The local heat transfer on a particular surface is given by $q = \rhoavg \cp \alpha \, \Nu \, (T_w - T_c)/\ell$ \citep{holman2002heat}, and insertion of the $\Nu$-$\Ra$ relationship gives
\begin{equation}
\label{wallfluxEq}
\int_S q \, dS  = \rhoavg \cp \left( \frac{g \alpha^2}{\nu \Tavg} \right)^{1/3} \int_S \cw \, (T_w - T_c)^{4/3} \, dS \, .
\end{equation}
Here, we have adopted the convention $X^p = \abs{X}^p \sgn{X}$ for any variable $X$ in order to shorten expressions. We will take the wall temperature, $T_w$, to be a different constant on each surface --- the ceiling, the floor, and the four sidewalls --- so that the integral in Eq.~(\ref{wallfluxEq}) reduces to a summation over these six faces. We note that the lengthscale $\ell$ has dropped out of Eq.~(\ref{wallfluxEq}) as a consequence of the $\Nu \sim \Ra^{1/3}$ law. Though easily overlooked, this cancellation simplifies the analysis considerably and avoids certain ambiguities in the definition of $\ell$ (see section S1 in the supporting information). This is one factor that will ultimately allow us to obtain {\em closed-form} expressions for the ventilation rate.

Both $T_c$ and $U$ are still unknown at this stage, and so a second relationship is required to close the model. Because the Dragon's Belly is a {\em single}-opening cave, we can leverage fundamental results from the study of two-layer exchange flows \citep{farmer1986maximal, dalziel1991two, helfrich1995time, pratt2008critical}. Such flows have been found to spontaneously organize into a critical state, characterized in this context by a composite Froude number equal to one. Imposing this condition gives the speed of the exchange flow as
\begin{equation}
\label{Ueq}
U = \sqrt{ \frac{1}{2} \frac{H g |T_e-T_c|}{\Tavg}} \,\, ,
\end{equation}
We note that other closure conditions could be used for caves with multiple openings \citep{christoforou1996air, flynn2006natural}.

\subsection{Estimating internal-surface temperatures}
\label{SectionTemps}
Our main goal is to solve Eqs.~(\ref{caveheatEq})--(\ref{Ueq}) for the cave temperature, $T_c$, and the exchange speed, $U$, but first it is necessary to estimate the cave-wall temperatures, $T_w$, for insertion into Eq.~(\ref{wallfluxEq}). Here, we will use temperature measurements from inside the cave to guide and validate certain approximations made in our model. Once constructed, though, the model will not rely on these internal measurements, and the same framework could be applied to other caves without any recalibration. 

We begin with a formula, often used in the design of wine cellars, to estimate the underground temperature at a depth $z$ \citep{pinchover2005introduction, tinti2014experimental},
\begin{equation}
\label{winecellar}
T(z,t) = \Tavg + \Delta T e^{-z/z_0} \cos \left(\omega (t-t_0) - z/z_0 \right)
\end{equation}
Here, $\omega = 1.99 \times 10^{-7}$~rad/s is the annual frequency, $\Delta T = 9.3$~K is the amplitude of the seasonal temperature variation (illustrated by the dashed curve in Fig.~\ref{Fig2}d), $\DiffLim = (5.6 \pm 1.6) \times 10^{-7}$~m$^2$/s is the thermal diffusivity of the limestone medium, $z_0=\sqrt{2 \DiffLim/ \omega} = 2.4 \pm 1.3$~m is the attenuation depth, and $t_0$ simply sets an initial time. This formula neglects higher modes, for example diurnal and synoptic variations, since they attenuate much more rapidly with depth. The attenuation depth, $z_0$, depends on the physical properties of the local geology, thus, by using appropriate values for thermal diffusivity temperature of surfaces of caves formed in other than limestone can be calculated.

Equation (\ref{winecellar}) is an exact solution to a diffusion problem in a homogeneous domain with periodically driven upper surface. Although the presence of the cave violates the homogeneity assumption, the much lower density of the enclosed air implies that a very large change in air temperature would be required to modify the limestone temperature appreciably. We therefore apply the formula to estimate the temperature of each cave surface. For the sidewalls, we estimate an average temperature by integrating the formula over the height of the wall with a 3-point trapezoid rule. Figure~\ref{Fig2}c shows the temperature range predicted by the wine-cellar formula as it varies with depth, note that at approximately 12~m depth, the temperature of the ground becomes equal to the annual mean of the outside temperature and there are no seasonal variations below 12~m. That 12~m depth depends on the physical properties of the local geology and varies depending on where the cave is formed. Notably, the temperature of the Dragon's Belly ceiling varies over a much wider range than does the floor (by a factor of 12).

We now compare these predictions with temperature measurements taken inside the cave. Internal convective flows tend to homogenize the cave-air temperature and so, regardless of where the sensors are placed, the measurements are influenced by all of the surrounding surfaces. We therefore define an average wall temperature, $\Twavg$, which incorporates the temperatures of the ceiling, floor, and sidewalls as weighted by surface area
\begin{equation}
\label{Twavg}
\Twavg = \frac{1}{S} \sum_{i=1}^6 S_i T_i \, .
\end{equation}
Here, $T_i$ and $S_i$ are the temperature and area of surface $i$, and $S = 1620$ m$^2$ is the total surface area. Figure~\ref{Fig2}d shows the estimated $\Twavg$ (black) as it compares with measurements (red). The prediction captures the phase of the measurements very accurately and under-predicts the amplitude slightly. The under-predicted amplitude is of minor consequence when compared to the scale of the {\em outside-temperature} fluctuations, shown by the faint gray data in Figure~\ref{Fig2}d. Due to their larger magnitude, the {\em outside-temperature} fluctuations are primarily responsible for creating the inside/outside temperature difference that drives exchange.

Figure~\ref{Fig2}d provides one last insight. The predicted $\Twavg$ agrees well with internal measurements, even though it does not take ventilation into account. This suggests that the cave's thermal exchange is a {\em wall-dominated process} --- a fact that is bourn out by scaling analysis in section S3 in the supporting information. It may seem paradoxical that, while ventilation is the main quantity we aim to predict, observations suggest it to be a secondary effect. The resolution is that, while ventilation is indeed secondary for {\em thermal} exchange, it is the {\em primary} transport mechanism for cave gases such as carbon dioxide and radon. By using simple internal-temperature estimates to predict ventilation rates, our model will be able to accurately describe the transport of these gases.

\subsection{Predicting exchange rates}
Having obtained reliable wall-temperature estimates, it is now possible to solve Eqs.~(\ref{caveheatEq})--(\ref{Ueq}) for the cave air temperature, $T_c$, and the speed of the exchange flow, $U$. Once $U$ is known, the main quantity we seek is the {\em ventilation rate}
\begin{equation}
\label{Ventilation}
\lambda_v = AU/V \, ,
\end{equation}
which will allow us to determine how cave gases are exchanged with the outside.

We solve Eqs.~(\ref{caveheatEq})--(\ref{Ueq}) with two different approaches. The first is a numerical strategy that tracks the temperature of each individual cave surface and dynamically switches thermal convection on or off depending on the direction of the local temperature gradient. This model thus represents the complex physics of thermal convection with spatial and temporal resolution. See section S2 in the supporting information for further details. The second approach is a course-grained model that assumes spatially uniform convection in order to arrive at a system amenable to perturbation analysis. This analysis ultimately gives an {\em explicit} expression for how the ventilation rate depends on cave and climatic parameters. See section S3 in the supporting information for details. With these two approaches, we have both the physical fidelity offered by a computational model and the transparency offered by a purely analytical model. Comparison between the two will allow us to test the additional assumptions made in the perturbation model.
	
As described in section S3 in the supporting information, analysis of the relevant physical scales identifies two dimensionless parameters, $\pi_1$ and $\pi_2$. Briefly, $\pi_1$ represents the timescale of wall-thermal exchange compared to the timescale of temperature variation (i.e.~a year), while $\pi_2$ represents the relative strength of ventilation-induced versus wall-induced thermal exchange. In terms of these parameters, perturbation analysis gives the ventilation rate as
\begin{equation}
\label{VentSoln}
\lambda_v \sim 
\frac{A}{V} \sqrt{ \frac{Hg }{2} } \,\,\, { \abs{ \frac{\Twavg - T_e }{\Tavg}} }^{1/2} \,\,
\left( 1 - 0.5 \left( \frac{\pi_2}{\chi_0} \right)^{3/4} \abs{ \frac{\Twavg - T_e}{\Tavg} }^{1/8} \right)
\, , \qquad \text{for } \pi_2 \ll 1 \, ,
\end{equation}
This formula shows that the ventilation rate depends primarily on the absolute difference between the cave-wall temperature, $\Twavg$, and the outside temperature, $T_e$. The term involving $\pi_2$ represents nonlinear damping of ventilation due to the internal air adjusting to the outside temperature. For the Dragon's Belly Cave, this term only makes a $1\%$ contribution. However, caves with less idealized features will see the contribution of this term, and future studies on such caves could allow us to examine it more closely.

Once $\lambda_v$ is known (from either the numerical or perturbation model), we can determine how cave gases --- namely carbon dioxide and radon-222 --- are exchanged with the outside environment. The transport of either quantity can be described by the ODE
\begin{equation} 
\label{CEq}
\dot{C} = \frac{S}{V} \source - \lambda_v(C - C_e) - \lambda_r C \, ,
\end{equation}
where $C$ is the concentration of {\CO} [ppmv] or radon [Bq/m$^3$] inside the cave. The first term on the right represents the flux of {\CO} or radon into the cave through internal surfaces. We assume the production term, $\source$, to be constant and will estimate its value from measurements. The second term on the the right represents exchange with the outside via ventilation. Here, $C_e$ is the external concentration, where $C_e = 392$ ppmv for {\CO} and $C_e$ is negligible for radon. The last term on the right represents decay of radon-222, where $\lambda_r = 2.1 \times 10^{-6}$ s$^{-1}$ (this term is zero for {\CO}).

\section{Results: Field measurments and theoretical predictions}
\begin{figure}
\includegraphics[width=1\textwidth,left]{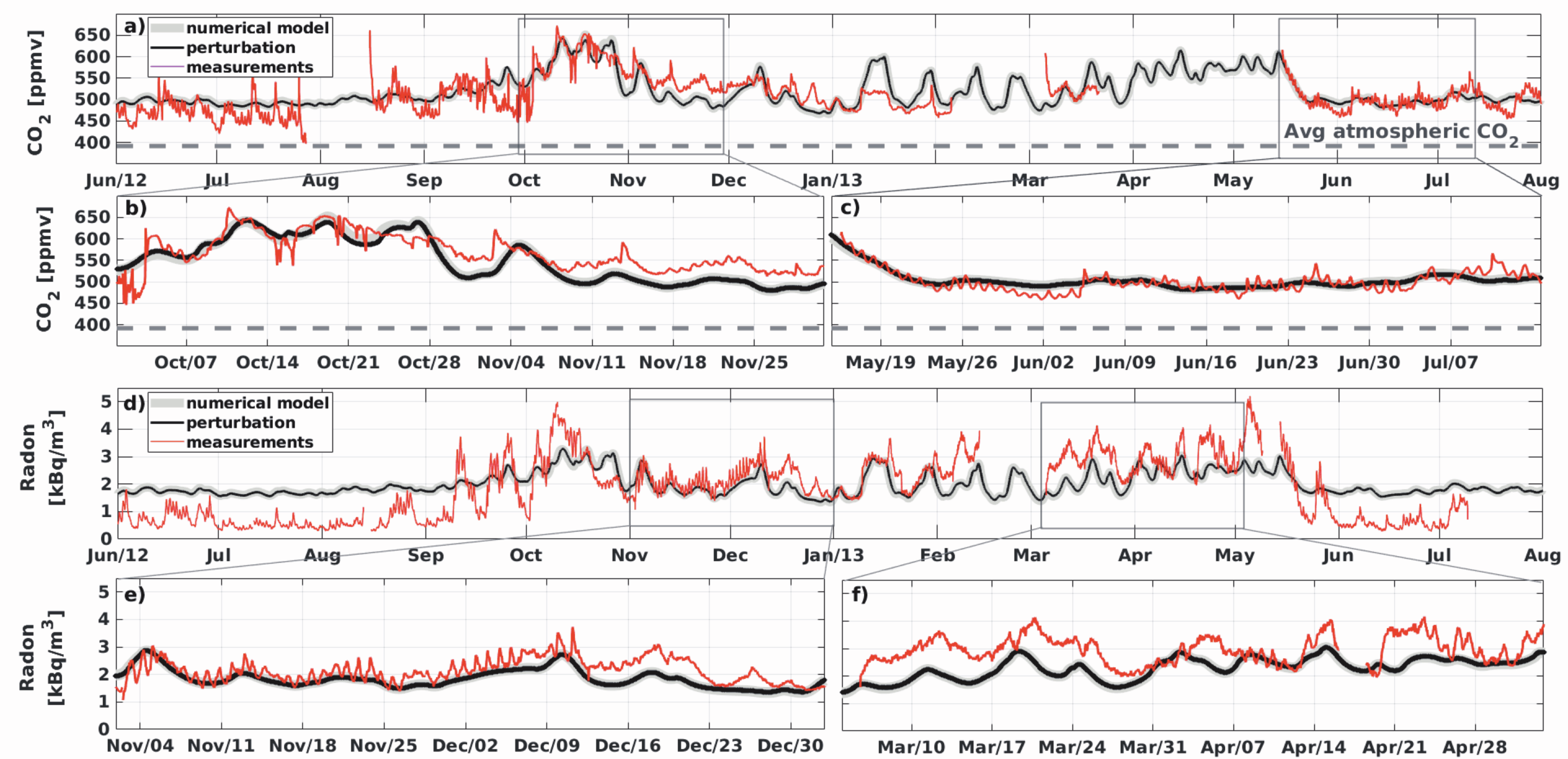}
\caption{Variation of in-cave gases. (a) The measured levels of {\CO} (red) over the full 14-month period, along with predicted values from the numerical model (gray) and the perturbation solution (black), the two of which are nearly identical. (b)--(c) Zoom of selected date ranges offers a closer comparison. (d)--(f) Similar plots for radon. Notice both gases show prominent variations on a near-weekly timescale, and the theory exhibits similar fluctuations.}
\label{Fig3}
\end{figure}

We now discuss the measurements taken at the Dragon's Belly site, shown by the red curves in Figs.~\ref{Fig3}~a--c for {\CO} and d--f for radon. Cave air gas concentrations in the Belly were collected at 30-minute intervals over a 14-month period, with occasional interruptions due to flooding and/or power outages. Both the perturbation model and the numerical model were performed at 24-hr resolution, models and measured data are plotted together in Fig. 3. The highest concentrations of cave gases occur during fall and spring, while the lowest concentrations occur during summer. This result is counter to those in caves under predominantly chimney-driven airflow regimes that promote strong winter-time ventilation (e.g. Obir Cave \citep{spotl2005cave}; Inner Space Caverns \citep{banner2007seasonal}; Hollow Ridge Cave \citep{kowalczk2010cave}).

During the winter, carbon-dioxide levels in the Belly are near their minimum, while radon levels are intermediate. Qualitatively, these trends suggest ventilation to be lowest during fall and spring, when daily average inside-outside temperature differences are small, and highest during summer and winter, when the temperature differences are maximal. The observations therefore support the idea of ventilation being caused primarily by buoyancy-driven flows.

Predictions of {\CO} and radon are shown by the dark curves in Fig.~\ref{Fig3}. The gray curves represent the numerical model, with convection switched on/off for each surface individually, while the black curves show the perturbation solution given by Eq.~(\ref{VentSoln}). The two predictions are nearly indistinguishable, providing validation for the additional assumptions made in the perturbation model. We estimate the source term, $\source$, by minimizing the difference between model and measurements, giving $\source = 3.1 \times 10^{-9}$~kg/(m$^2$s) for {\CO} and $\source = 25 \times 10^{-6}$ kBq/(m$^2$s) for $^{222}$Rn. We note that our estimated source terms for radon and {\CO} is comparable to others found in the literature (e.g. Radon: $\source = 9.9 \times 10^{-6}$ kBq/(m$^2$s) \citep{faimon2006anthropogenic}; $\source = 13~61 \times 10^{-6}$ kBq/(m$^2$s) \citep{kowalczk2010cave}; {\CO}: $\source~=~1.989~\times~10^{-10}$~kg/(m$^2$s) \citep{baldini2006spatial}; $\source~=~1.56~\times~10^{-9}$~kg/(m$^2$s), \citep{faimon2006anthropogenic}). With the source term as the only fit parameter, both the {\CO} and radon predictions correspond well with  measurements over the majority of the 14-month study. In particular, the model accounts for the higher levels of cave gases in the fall/spring and lower levels in the summer/winter. The only obvious discrepancy is in comparing radon levels in the summer, where the theory predicts a significantly higher level than was measured. Carbon-dioxide shows good agreement during this same period, though, suggesting the discrepancy to be specific to radon. 

While not a focus of this study, hydrologic conditions within the rock above the Belly may have a significant influence on agreement between modeled and measured $^{222}$Rn over seasonal time periods. \cite{andrews1972mechanism} demonstrated that radon emanation from limestone occurs from within a surface layer only a few microns thick, and that under dry conditions, alpha-recoil causes radon gas to be ejected from the crystal lattice into inter-crystal cracks and imperfections, eventually to find its way out of the limestone through relatively slow diffusion in air at approximately 10$^{-2}$ cm$^2$ s$^{-1}$. Conversely, under wet conditions within the epikarst, alpha-recoil can cause radon particles to be ejected directly into pore water. As dripwater percolates downward, it has a scavenging effect, driving a significantly higher flux of radon during periods when the epikarst is hydrologically saturated. An excellent example of this phenomenon was illustrated by \cite{kowalczk2010cave} at Hollow Ridge Cave, when a tropical storm delivered 83~mm of rainfall, saturating the epikarst and causing a nearby river to flood and plug up the lower entrances of the cave, temporarily ceasing ventilation. Directly following this rainfall, $^{222}$Rn emanation rates increased from 'normal' 48-222~Bq~m$^{-2}$~hr$^{-1}$ to approximately 1200~Bq~m$^{-2}$~hr$^{-1}$. Unpublished data from laboratory testing of limestone from both Dragon's Tooth and Hollow Ridge Cave confirms that wet transport can increase $^{222}$Rn emanation by 1.6$\sim$2.4 times dry transport (William Burnett - FSU, personal communication, 2015).  

\begin{figure}
\includegraphics[width=1\textwidth,left]{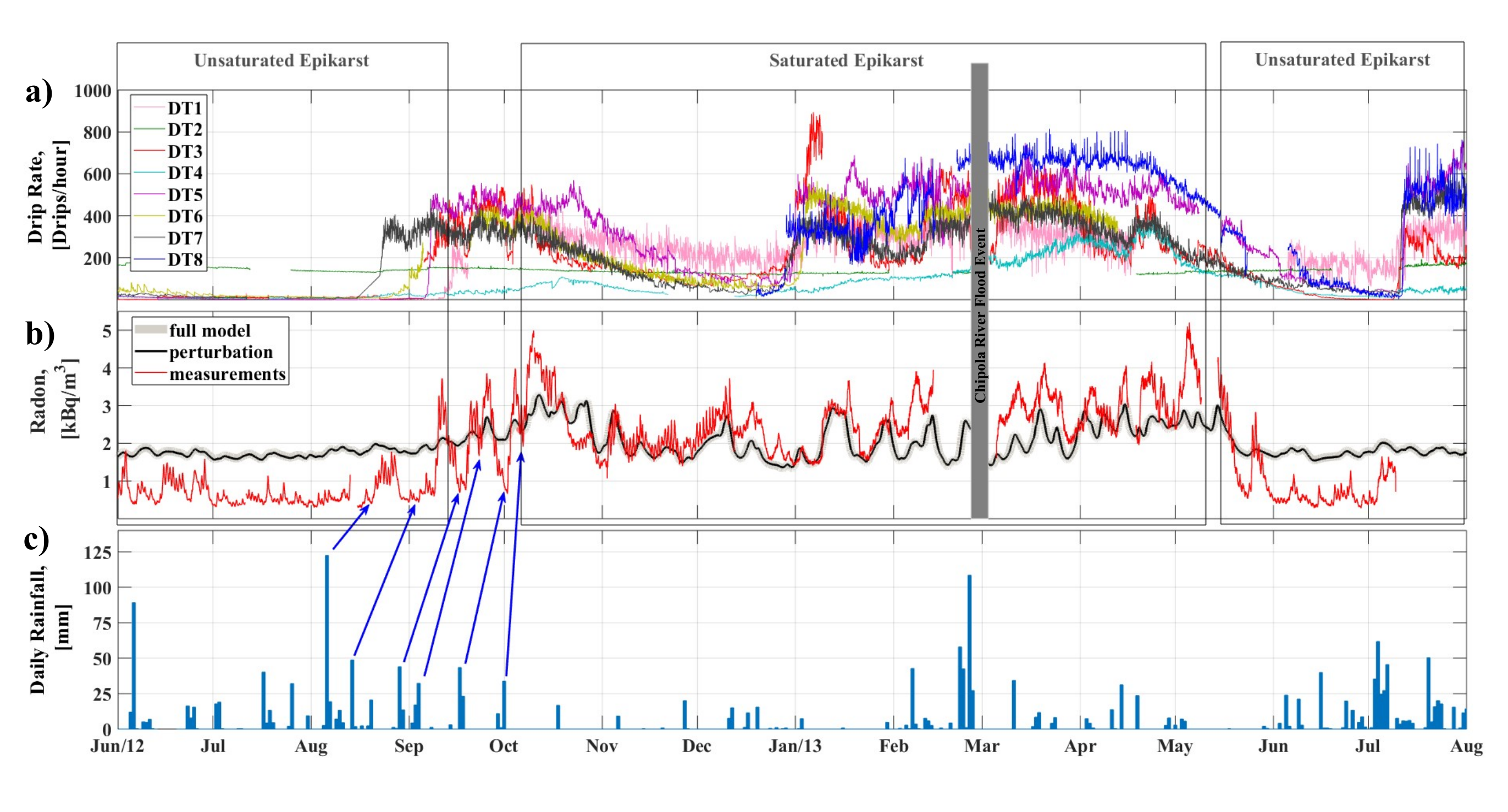}
\caption{Hydrological conditions within the rock above Dragon's Belly. (a) The measured drip rates at eight locations with Dragon's Belly. (b) The meadured levels of radon (red) over the full 14-month period, along with predicted values from the numerical model (grey) and the perturbation solution (black), (c) Daily rainfall above Dragon's Belly. Blue arrows indicate radon response to several successive rainfalls.}
\label{Fig4}
\end{figure}

Figure \ref{Fig4} illustrates that prior to August 2012, most of the monitored drip sites were essentially zero. During August, a 120~mm rainfall and several successive $\sim$40~mm rainfalls prompted hydroligcal saturation within the overburden. Approximately 1-2 weeks after each event, an increase in $^{222}$Rn was observed, followed by a decay to background levels. In September 2012, all monitored sites rapidly increased in drip rate as the epikarst came to full saturation. These saturated conditions persisted until May 2013, when lower rainfall and increased evapotranspiration conspired to reduce the hydrologic saturation, causing measured radon to fall below values predicted by our model.

In additional to the seasonal variation, the {\CO} and radon measurements show more rapid fluctuations on a few characteristic scales. Figures \ref{Fig3} b--c and e--f zoom on selected date ranges to highlight these. First, the measurements show very rapid, diurnal fluctuations, which result from the different day/night temperatures. In the model, we only input {\em daily} outside temperatures, and so these very fast fluctuations are not present. Presumably they could be captured with more temporal resolution though (e.g.~hourly readings). Interestingly, there is a second, intermediate timescale --- on the order of 5--9 days --- that is also evident in the {\CO}/radon data. This nearly weekly cycle is likely caused by synoptic-scale meteorology \citep{holton2012introduction}, though manmade activity may play a role too \citep{daniel2012identifying, earl2016weekly}. Regardless, the model shows variations over exactly the same scale. This implies that the weekly cycle in cave-gas levels is ultimately tied to a weekly cycle in outside temperatures that drives an exchange flow as described by our theory. Over certain periods, for example Fig.~\ref{Fig3}e, the theory even shows quantitative agreement with tracer measurements.

\section{Conclusions}
Here, we have constructed a theoretical framework to predict subterranean-cave ventilation rates from a minimal set of external information --- namely, knowledge of the outside temperature and the physical dimensions of the cave. The major advantage is that external temperatures are much more easily measured or, in the application of climate reconstruction, estimated, than are conditions inside the cave. Through scaling analysis and a few key modeling assumptions, we obtained explicit relationships for how ventilation depends on system parameters, given by Eq.~(\ref{VentSoln}). Comparison with time-resolved, in-situ measurements demonstrates the theory's ability to accurately describe seasonal and synoptic-scale fluctuations of transported cave gases. As an immediate application, this theory could be used to improve estimates for production rates of cave gases such as CO$_2$ and $^{222}$Rd. Currently-used field techniques can be destructive to the cave's fragile environment, but our methodology is completely non-invasive as it relies only on knowledge of outside temperatures.

A longer-term application is to use these results to improve speleothem interpretation. Inferring past climate from speleothems is inherently an inverse problem, typically treated by representing the climate's influence on the proxy (e.g.~speleothem, tree ring, etc.) via a transfer function, which is then inverted. While it has been recognized that ventilation plays a critical role in speleothem growth, such effects have not yet been quantitatively incorporated into paleoclimate inference models, likely due to the complexity involved. Our model, however, offers explicit formulas for how ventilation depends on cave and climatic variables. The transparency afforded by this model could therefore prove useful in obtaining tractable transfer functions for climate reconstruction. In essence, the simpler the forward model, the more feasible inference becomes.

\section{Acknowledgments}
Geophysical Fluid Dynamics Institute contribution number xxx. This research was supported by the NSF Grant AGS-1032403; additional support for K. Khazmutdinova was provided through teaching and research assistantships at Florida State University. The authors sincerely thank the management staff of Florida Caverns State Park, especially Kelly
Banta, for providing an incredible opportunity to study the Dragon's Tooth Cave. Thanks to Cameron Ridgewell for creating a 3D map of the Dragon's Tooth Cave. We thank William Burnett for help with interpreting radon measurements, and William Dewar for inspiring suggestions and comments. The meteorological data from Marianna Airport is obtained from \url{https://www.ncdc.noaa.gov/cdo-web/}. The input files and results of the numerical calculations are available from the authors upon request (kk11m@my.fsu.edu).
\bibliographystyle{plain}
\bibliography{bibliography.bib}
\end{document}